\documentstyle[epsfig]{mn}

\def\spacing{\renewcommand{\baselinestretch}{1.0}}
\def\lessapprox{\,\raise 0.6ex\hbox{$<$}\kern -0.75em\lower 0.47ex
    \hbox{$\sim$}\,}
\def\largapprox{\,\raise 0.6ex\hbox{$>$}\kern -0.75em\lower 0.47ex
    \hbox{$\sim$}\,}

\def\wth{\omega(\theta)}

\def\ol{\Omega_\Lambda}
\def\om{\Omega_m}
\def\be{\begin{equation}}
\def\ee{\end{equation}}
\def\hth{\hat{\phi}}

\spacing

\begin{document}
\title[Weak Lensing and $\wth$]
{The Effect of Weak Lensing on the Angular Correlation
Function of Faint Galaxies}
\author[R. Moessner, B. Jain and J. V. Villumsen]
{R. Moessner, B. Jain and J. V. Villumsen\\
Max Planck Institut f\"{u}r Astrophysik\\
Karl Schwarzschild-Str. 1,\\
85740 Garching, Germany\\
}
\maketitle
\begin{abstract}
\noindent

The angular correlation function  $\wth$ of faint galaxies is affected
both by nonlinear gravitational evolution and by magnification bias due to
gravitational lensing. We compute the resulting $\wth$
for different cosmological models and show how its shape and redshift
evolution depend on $\Omega$ and $\Lambda$.
For galaxies at redshift greater than 1 ($R$ magnitude fainter
than about 24), magnification bias can significantly enhance
or suppress $\wth$, depending on the slope of the number-magnitude
relation. We show for example how it changes the ratio of $\wth$
for two galaxy samples with different number-count slopes.
\end{abstract}


\section{Introduction}

The angular correlation function of galaxies has been used to
characterize the large-scale distribution of galaxies for over
2 decades. If the number density on the sky at angular position
$\hat{\psi}$ is $n(\hat{\psi})$, then $\wth$ is defined as
\begin{equation}
\wth=\frac{\langle n(\hat{\psi})n(\hat{\phi})\rangle}{\bar{n}^2}-1 \; .
\label{wthav}
\end{equation}
The  3-dimensional unit vectors $\hat{\psi}$ and $\hat{\phi}$
are used to define the angular separation $\theta$ as
$ \hat{\psi} \cdot \hat{\phi}  = \cos(\theta)$, and $\bar{n}$ is the
mean galaxy number density on the sky.  The observed
galaxy distribution is well described by a
power law $\wth\propto \theta^{-\gamma}$, with slope
$\gamma\simeq 0.8$. Since $\wth$ is a projection on the sky of the
3-dimensional auto-correlation function $\xi(r, z)$, the above power
law for $\wth$ has been associated with a power law for $\xi$ with
slope $1.8$ which is close to the observed value in the nearby galaxy
distribution.
Measurements of $\wth$ are difficult to interpret as it involves a
projection of the galaxy distribution out to redshifts of order 1
(e.g. for a sample with limiting $R$ magnitude of about 24.5). Thus the effects
of evolution of large scale structure due to gravitational clustering
as well as galaxy evolution need to be understood to interpret $\wth$.

Recently Villumsen (1996) has considered the effect on $\wth$ of
gravitational lensing by large-scale
structure along the line of sight (Gunn 1967).
Lensing increases the area of a given patch on the sky, thus diluting
the number density. On the other hand, galaxies too faint to be
included in a sample of given limiting magnitude are brightened due
to lensing and may therefore be included in the sample. The net
effect, known as magnification bias,
can go either way: it can lead to an enhancement or suppression of the
observed number density of galaxies, depending on the slope of the
number-magnitude relation.  Variations in the number density which
are correlated over some angular separation alter $\wth$.
Villumsen (1996) showed how the linear evolution of density fluctuations
along the line of sight can be used to compute the change in $\wth$
due to magnification bias.
The deflections of neighbouring photon trajectories
due to lensing by large scale structure are very small, hence the
calculation can be done in the limit of weak lensing.

In this paper we compute $\wth$ for different cosmological models
taking into account the effects of nonlinear
gravitational evolution and gravitational lensing.
Since $\wth$ is a 2-point statistic, even in the fully nonlinear
regime it is determined completely by the 3-dimensional power spectrum.
We use extensions of the proposal of Hamilton et al. (1991) to
include the nonlinear evolution of the power spectrum into the
calculation of $\wth$. We also include the dependence on
the cosmological parameters $\Omega_m$ and $\Omega_{\Lambda}$.

In Section 2 the formalism for computing $\wth$ is presented.
Section 2.1 discusses the effects of the cosmological model via
the growth of density perturbations and the distance-redshift relation.
Results for CDM-like power spectra for different values of
$\Omega_m$ and $\Omega_{\Lambda}$ are presented in Section 3.
We explore ways to isolate
the effect of the lensing contribution to $\wth$ in Section 4 and
conclude in Section 5.

\section{The effect of magnification bias on the angular correlation function}

The effect of magnification bias due to weak gravitatinal lensing on
the galaxy angular correlation function $\wth$
has been derived in Villumsen (1996).
Written in terms of the time-dependent power spectrum instead of the
present day one, which makes it applicable to nonlinear
power spectra, $\wth$ is given by
\begin{eqnarray}
\wth&=&4\pi^2\int_0^{\chi_H} d\chi\ \left[
W(\chi)a(\chi)b(\chi)+15\Omega_m (s-0.4)g(\chi)\right]^2 \nonumber \\
&&\times \int_0^\infty dk\, k\, {P(\chi, k)\over a^2(\chi)}\,
J_0\left[k r(\chi)\theta\right]\, .
\label{wfull}
\end{eqnarray}
In this section the physical meaning of the various
terms in the above equation and the notation used will be clarified.

We have used the notation of Jain \& Seljak (1997), where
$\chi$ is the radial
comoving distance, $\chi_H$ that to the horizon, $r(\chi)$ is the
comoving angular diameter distance, $W(\chi)$ the normalized
radial distribution of galaxies in the sample, and
\begin{equation}
g(\chi)= r(\chi) \int_\chi^{\chi_H}
{r(\chi' -\chi) \over r(\chi')}W(\chi')d\chi'\ .
\end{equation}
{}From the expression for the unperturbed line element
\begin{equation}
ds^2=a^2(\tau)\left( -d\tau^2+ d\chi^2+r^2(d\theta^2+\sin^2 \theta d\phi^2)
 \right)  ,
\end{equation}
$\tau$ being conformal time, it follows that
\begin{eqnarray}
r(\chi)=\sin_K\chi \equiv
\left\{ \begin{array}{ll} K^{-1/2}\sin K^{1/2}\chi,\ K>0\\
\chi, \ K=0\\
(-K)^{-1/2}\sinh (-K)^{1/2}\chi,\ K<0
\label{rchi}
\end{array}
\right.
\end{eqnarray}
where $K$ is the spatial curvature.
Note that for a delta-function distribution of galaxies,
$W(\chi')=\delta(\chi'-\chi_S)$, $g(\chi)$ reduces to
$g(\chi)=r(\chi)r(\chi_S-\chi)/r(\chi_S)$.
This notation is related to the one of Villumsen (1996) through
$\chi=x$, $r(\chi)=y(x)$, $W(\chi)/r^2(\chi)=S(x)$ and $g(\chi)/r(\chi)=w(x)$.

First we shall briefly describe the effect of magnification bias
on $\wth$. Gravitational lensing of a galaxy by dark matter concentrations
between us and the galaxy increases the area of the galaxy image while
conserving surface brightness. This results in a magnification $\mu$ which is
given by the ratio of lensed to unlensed area of the image. The
amplitude of $\mu$ depends on the convergence $\kappa$, which
is the surface mass density divided by the critical density,  and
the shear $\gamma$ through (Young 1981)
\be
\mu=\frac{1}{(1-\kappa)^2-\gamma^2}
\label{mugen}
\ee
In the limit of weak lensing, $|\kappa|, |\gamma| \ll 1$,  applicable
to lensing by large-scale structure, the above expression  reduces to
\be
\mu=1+2\kappa
\label{mu}
\ee
This magnification has two effects. Since the lensed area is
increased due to deflection of the light rays, the number density of
galaxies decreases inversely proportional
to $\mu$. There is a competing effect, however. In a flux limited
survey, magnification brings some faint galaxies  above the flux limit,
which would not otherwise have been detectable, thus increasing
the number density of galaxies.   Which of the effects wins depends on the
reservoir of faint galaxies available, which can be quantified by
the slope $s$ of the true number counts $N_0(m)$ for a magnitude limit
$m$,
\be
s= \frac{d \log N_0(m)}{dm} \; .
\ee
The two effects change the galaxy numbers by (e.g. Broadhurst, Taylor \&
Peacock 1995)
\be
N^\prime(m)=N_0(m) \mu^{2.5s-1} \,
\ee
which reduces to
\be
N^\prime(m)=N_0(m)\left[ 1+5(s-0.4)\kappa \right]
\label{Np}
\ee
in the weak lensing limit.
In addition, the number density of galaxies is changed by $\delta n$
due to intrinsic clustering.
Let $\bar{n}$ denote the average number density of galaxies. At a
given position  in the sky $\hth$, the number density is thus
changed to
\be
n(\hth)=\bar{n} \left( 1+ \delta n(\hth) +5(s-0.4)\kappa(\hth)
\right) \; .
\label{nth}
\ee
Now we further assume that galaxies trace the underlying dark matter
distribution so that the 3-dimensional galaxy overdensity is,
\be
\delta_g(\vec x) = b \ \delta(\vec x) \; ,
\ee
where $\delta(\vec x)$ is the dark matter overdensity and the bias $b$ is the
proportionality factor.
In this linear bias model, the perturbed, projected number density is
given by (Villumsen 1996)
\be
\delta n(\hth)= b \int_0^{\chi_H} d \chi W(\chi) \delta\left(r(\chi) \
\hth\right) \; .
\label{ndelta}
\ee
whereas the convergence $\kappa$ is given by (Villumsen 1996)
\be
\kappa(\hth)= \frac{3}{2} \om \int_0^{\chi_H} d \chi g(\chi)
   \frac{\delta(r(\chi) \hth)}{a}
 \; .
\label{kappa}
\ee
The angular correlation function $\wth$ as defined in
equation~\ref{wthav} involves the expectation
value $\langle n(\hth) n(\hat{\psi})\rangle $.
Since both $\delta n$ due to intrinsic clustering, and $\kappa$
are related to $\delta$ according to Eqs.~\ref{ndelta} and
{}~\ref{kappa}, intrinsic dark matter correlations will induce
correlations among galaxies. There are three distinct effects:
correlations among the galaxies tracing correlated dark matter densities,
described by
$\langle \delta n(\hth) \delta n(\hat{\psi}) \rangle$; among background
galaxies
being lensed by correlated dark matter concentrations, described by
$\langle \kappa(\hth) \kappa(\hat{\psi}) \rangle$; and finally among
background
galaxies and the foreground
galaxies tracing the correlated dark matter concentrations, which
are responsible for the lensing of the background galaxies, described
by the term $\langle \delta n(\hth) \kappa(\hat{\psi}) \rangle$.

The expressions for $n(\hth)$, $\kappa(\hth)$ and $\delta n(\hth)$
from the above equations are inserted into Eq.~\ref{wthav}. Expressing
$\delta(\vec x)$ in terms of its Fourier
transform $\delta(\vec k)$ and performing the ensemble average
$\langle \delta(\vec k) \delta(\vec k^\prime) \rangle$ then introduces
the power
spectrum $P(\chi,k)$. One finally obtains the result of
Eq.~\ref{wfull} (Villumsen 1996) by  evaluating the integrals,
under the further assumption
of $\theta \ll 1$, and using the plane-parallel approximation that only
Fourier modes with wave vectors nearly perpendicular to the line of
sight $\hth$ contribute to the integral.

In the previous paragraphs we discussed the origin of  the  three terms in
the angular correlation
function, obtained by multiplying out the squared bracket in Eq.~\ref{wfull}.
We can write them schematically as,
\begin{equation}
\wth= \omega_{gg}(\theta)+ \omega_{gl}(\theta)+\omega_{ll}(\theta)  \; .
\label{wterms}
\end{equation}
The third term $\omega_{ll}(\theta)$ is proportional to the
correlation function $C_{pp}(\theta)$ of galaxy image
ellipticities (Villumsen 1996),
\be
\omega_{ll}(\theta) = (2.5s-1)^2 C_{pp}(\theta) \; ,
\ee
The dependence of this term on $\om$ and $\ol$ is discussed in detail
in Jain \& Seljak (1997) (see also Bernardeau et al. 1997; Kaiser 1996).

\subsection{Dependence of $\wth$ on cosmological parameters}

With the assumption of constant linear bias, we can factor out the
constants involving bias, $s$ and $\Omega_m$,  and write $\wth$ as
\begin{eqnarray}
\wth&=&4 \pi^2 \, [\, b^2 \tilde{w}_{gg}(\theta)+ 30b\Omega_m(s-0.4)
\tilde{w}_{gl}(\theta) \nonumber \\
&& + 225\Omega_m^2(s-0.4)^2 \tilde{w}_{ll}(\theta)
] \; .
\label{wtilde}
\end{eqnarray}
Besides the explicit $\om$ dependence, $\wth$ depends on $\om$ and $\ol$
mainly  through the evolution of the power spectrum and the dependence
of $r(\chi)$ on $\Omega_m$ and $\Omega_{\Lambda}$.

In the linear regime the power spectrum depends on
the linear growing mode of density perturbations $D(\chi)$ and the
normalization $\sigma_8$ as,
\be
P(\chi,k) \propto \left[\sigma_8\, D(\chi)\right]^2 \; .
\ee
The linear growing mode is well approximated by (Carroll, Press \&
Turner 1992)
\begin{eqnarray}
D(\chi)&=&\frac{5}{2}\, a \, \Omega(a) \
[\Omega(a)^{4/7}- \lambda(a)  \nonumber \\
&&+ (1+\Omega(a)/2)(1+\lambda(a)/70) ]^{-1}  \; ,
\end{eqnarray}
where we have defined, following Mo, Jing \& Boerner (1997),
the time dependent fractions of density in matter and
vacuum energy, $\Omega(a)$ and $\lambda(a)$ in terms of the present-day
values $\om$ and $\ol$,
\be
\Omega(a)=\frac{\om}{a+\om(1-a)+\ol(a^3-a)}
\ee
and
\be
\lambda(a)=\frac{a^3\ol}{a+\om(1-a)+\ol(a^3-a)} \; .
\ee
The spatial
geometry also differs in different models, leading to a dependence of
the angular distance $r(\chi)$ on $\om$ and $\ol$ according to Eq.~\ref{rchi}.
These effects are contained in the terms $\tilde{w}_{gg}(\theta)$,
$\tilde{w}_{gl}(\theta)$ and  $\tilde{w}_{ll}(\theta)$ in Eq.~\ref{wtilde}.

On small scales (high-$k$) the growth of the power spectrum is
significantly affected by nonlinear gravitational clustering.
The nonlinear power spectrum is obtained from
the linear one through the fitting formulae of Jain, Mo \& White (1995) for
$\om=1$, and from those of Peacock \& Dodds (1996) for the open and $\Lambda-$
dominated models.
These fitting formulae are based on the idea of relating the
nonlinear power spectrum at scale $k$ to the linear power spectrum at
a larger scale $k_L$, where the relation $k(k_L, a)$ depends on
the power spectrum itself (Hamilton et al. 1991).
They have been calibrated from and tested extensively against N-body
simulations. For the linear power spectrum we take a CDM-like spectrum
(Bardeen et al. 1986) with shape parameter $\Gamma=0.25$, which
provides a reasonable fit to the observed galaxy power spectrum.
With this choice of the linear power spectrum, we use the fitting
formulae for the fully nonlinear power spectrum as a function
of $a$ and $k$ to evaluate  equation~\ref{wfull} for $\wth$.

Having specified the shape of the mass power spectrum, we need
to set the global cosmological parameters $\Omega_m$
and $\Omega_{\Lambda}$. The normalization of the power spectrum
in turn will depend on these parameters.
The four different cosmological models we consider include:
a flat universe with $\Omega_m=1$, an open
model with $\Omega_m=0.3$, and a flat $\Lambda-$
dominated model, with $\Omega_m=0.3$ and
$\Omega_{\Lambda}=0.7$. These three models are  normalized to cluster
abundances, which give the approximate relation
(White et al. 1993; Viana \& Liddle 1996; Eke, Cole \& Frenk 1996; Pen 1996):
\begin{equation}
\sigma_8= 0.6 \, \Omega_m^{-0.6} \; .
\end{equation}
(For $\Omega_m=0.3$, we use $\sigma_8= 1$, which is slightly lower
than given by the above relation.)
Our fourth model is  a flat universe, $\om=1$,  with a higher
normalization of  $\sigma_8=1$.

\section{Results for $\wth$ with CDM-like power spectra}
\label{secres}

While  the cosmological model is specified above,
the galaxy population still needs to be described in terms
of its redshift distribution,
the number count slope $s$ and the bias parameter $b$.
We model the redshift distribution of galaxies by
\begin{equation}
n(z)=\frac{\beta z^2}{z_0^3 \Gamma[3/\beta]} \exp(-(z/z_0)^{\beta})\; ,
\label{nz}
\end{equation}
for $\beta=2.5$, which agrees reasonably well with the redshift
distribution estimated
for the Hubble deep field from photometric redshifts (Mobasher et al.,
1996). $W(\chi)$ is related to the redshift distribution by
$W(\chi)=n(z)H(z)$, where the expressions for the Hubble constant $H(z)$
and the radial comoving distance $\chi(z)$ depend on the background cosmology.
The mean redshift is  given by
\begin{equation}
\langle z \rangle=\frac{\Gamma(4/\beta)}{\Gamma(3/\beta)} z_0  \; .
\end{equation}
For a flat model with $\Omega_m=1$, we obtain the
$R$ magnitude limit corresponding to a given mean
redshift by interpolation of a set of estimated values
from Charlot (1996).  In
this section we consider
red and blue galaxy populations, with number count slopes $s_r=0.25$ and
$s_b=0.45$ respectively,  and a fixed bias $b=1/\sigma_8$
for both samples. In the next section we will consider the more realistic
case of a relative bias for the different colour samples, since red
galaxies are more strongly clustered.

\begin{figure}
 \centerline{\psfig{figure=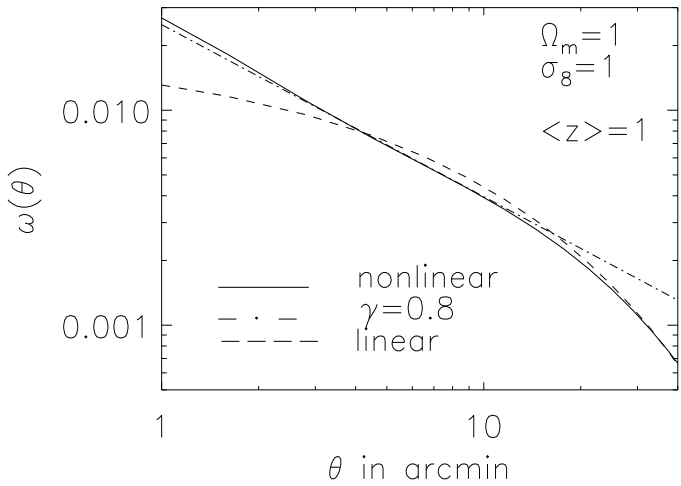,height=3in,width=4in}}
\spacing
 \caption{Angular correlation function as a function of angle for
	  a mean redshift of $1$, $\Omega_m=1$, $\sigma_8=1$ and
	  fixed bias $b=1$. The slope of the number counts relation
	  is $s=0.4$. The dashed line shows the linear
 prediction, while the solid line is the nonlinear prediction. The
nonlinear $\wth$ is closer to the power law with slope $-0.8$ shown
by the dash-dotted line.
	   }
 \label{wtheta1}
\end{figure}

Figure~\ref{wtheta1} shows $\wth$ for the $\om=1$, $\sigma_8=1$ model.
The linear and nonlinear predictions, shown by the solid and
dashed curve, are compared with a power law shape with slope $\gamma=0.8$.
These curves are for $s=0.4$, for which the lensing term drops out,
and therefore show the contribution of
the clustering term to $\wth$. The effect of nonlinear evolution
enhances $\wth$ for $\theta<4'$, while it slightly suppresses $\wth$
for $4'<\theta<20'$. On larger scales the linear and nonlinear
curves coincide. It is interesting that the nonlinear $\wth$ is much
closer to the observed power law than the linear curve which reflects
the curvature of the CDM power spectrum. At large angles, $\theta>10'$
the $\wth$
curve starts to deviate from the power law shape. For the open and $\Lambda$
dominated models, the results are similar, with the nonlinear
contribution being larger than the $\om=1$ model.

\begin{figure}
 \centerline{\psfig{figure=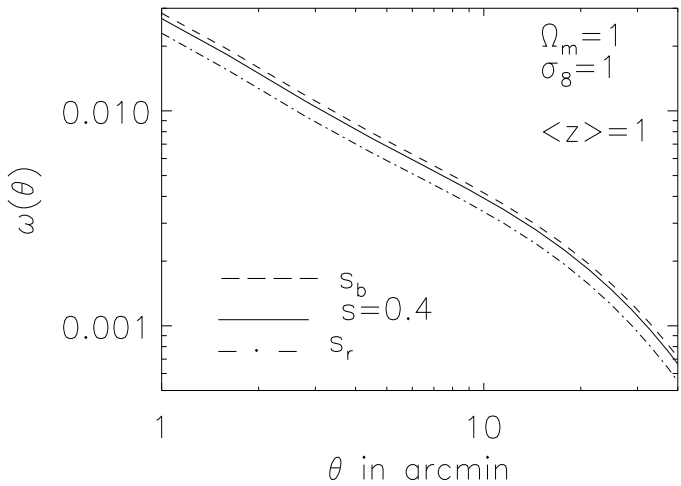,height=3in,width=4in}}
\spacing
 \caption{Angular correlation function as a function of angle for
	  three values of the slope of the number counts relations:
	  $s=0.25, 0.4, 0.45$. As in Fig.~\ref{wtheta1}, the
	  mean redshift is $1$, $\Omega_m=1$, $\sigma_8=1$ and
	  the bias is $b=1$.}
\label{wtheta}
\end{figure}

In Fig.~\ref{wtheta} we show how lensing affects $\wth$
in the flat $\Omega_m=1$ cosmological model, for three values of the
number count slope  $s=s_b=0.45$, $s=s_r=0.25$, and $s=0.4$.
The  mean redshift is~$1$, corresponding to an $R$ magnitude
limit of about $24.5$. The $s=0.4$ curve
represents the intrinsic clustering as in the previous figure.
For  $s_b>0.4$, the  amplitude
of $\wth$ is increased due to lensing, and for $s_r<0.4$ it is decreased.
Magnification bias hardly changes the shape of $\wth$ since the
three curves are nearly parallel to each other.

\begin{figure}
 \centerline{\psfig{figure=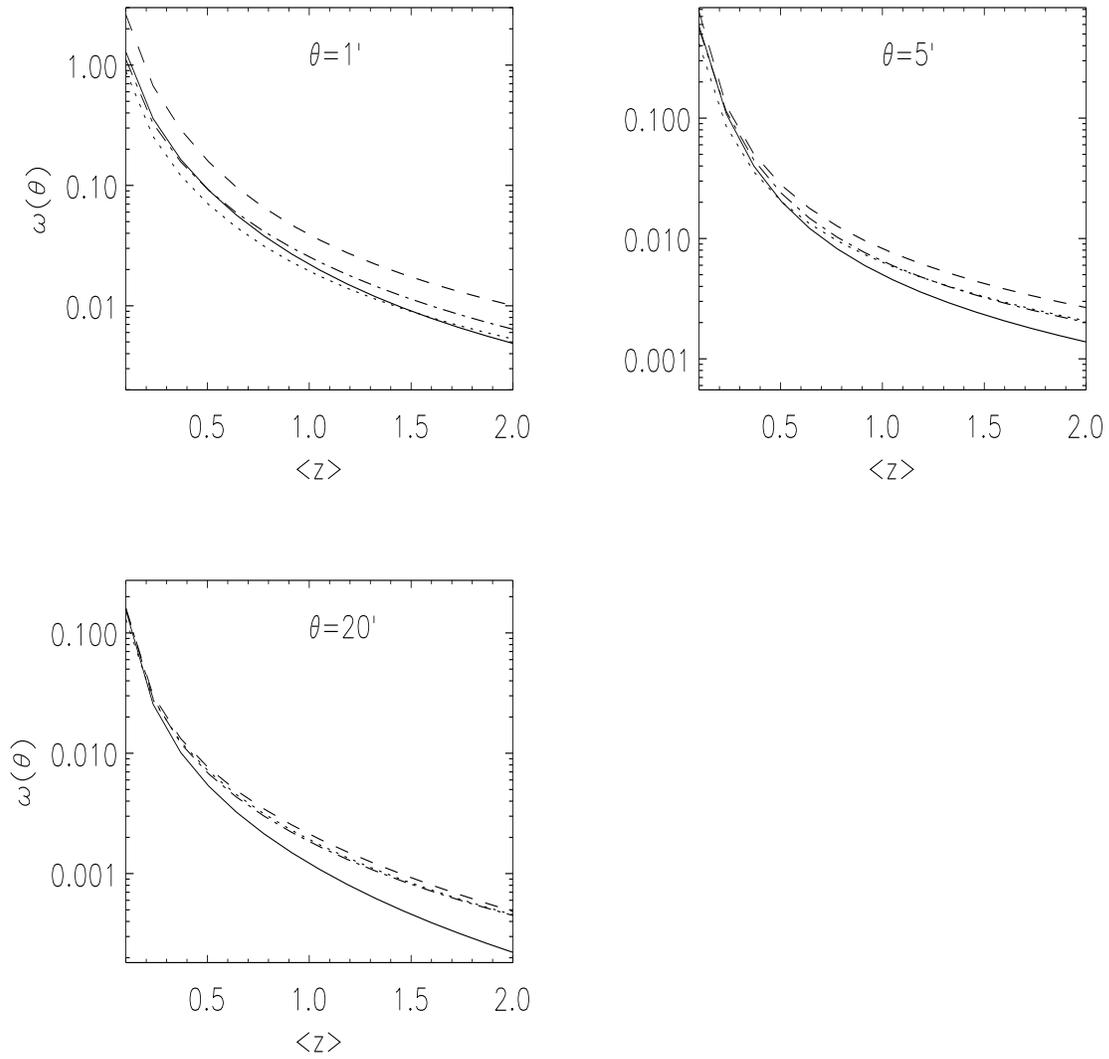,height=6in,width=6in}}
\spacing
 \caption{Comparison of angular correlation function due to intrinsic
	  clustering ($s=0.4$) for the four models, as a function of
	  mean redshift of the sample. Plots for three angles are
	   shown. The bias is fixed as $b=1/\sigma_8$; the four models
 are:  dash-dotted line: $\om=1$, $\sigma_8=1$; dotted line:
$\om=1$, $\sigma_8=0.6$; dashed line:  $\om=0.3$, $\sigma_8=1$; solid
line: $\om=0.3$, $\ol=0.7$,$\sigma_8=1$.}
 \label{wintrinsic}
\end{figure}

In Fig.~\ref{wintrinsic} we show the dependence of $\wth$ on
mean redshift of the sample. The contribution due to
intrinsic clustering is shown
for the four cosmological models, i.e. setting
$s=0.4$. Three values of $\theta$ are shown, ranging from
$\theta=1^\prime$, where nonlinear effects are important, to
$\theta=20^\prime$
where nonlinear effects become unimportant.
At $\theta=1^\prime$, the models with low $\om$ have higher amplitudes than
the models with $\om=1$. This is because in the former models the
growth of perturbations is slowed down at late times; when normalizing
to present-day cluster abundances, this implies a higher
normalization at earlier times compared to the $\om=1$ models. This
in turn means that nonlinear effects become important earlier on,
and lead to a greater effect of nonlinear clustering by today.
At large mean redshifts, however, the amplitude for the
$\Lambda-$model dips below the $\om=1$, $\sigma_8=1$ model. Here
a competing geometrical effect starts to come in: at a given redshift
a physical volume element is larger if  $\Lambda>0$ than in the
absence of
a cosmological constant, and therefore the intrinsic clustering is decreased.
For $\theta=20^\prime$, this geometrical effect in the $\Lambda-$model
is more noticeable and starts to be seen at lower mean redshifts
because competing nonlinear effects are unimportant.

\begin{figure}
 \centerline{\psfig{figure=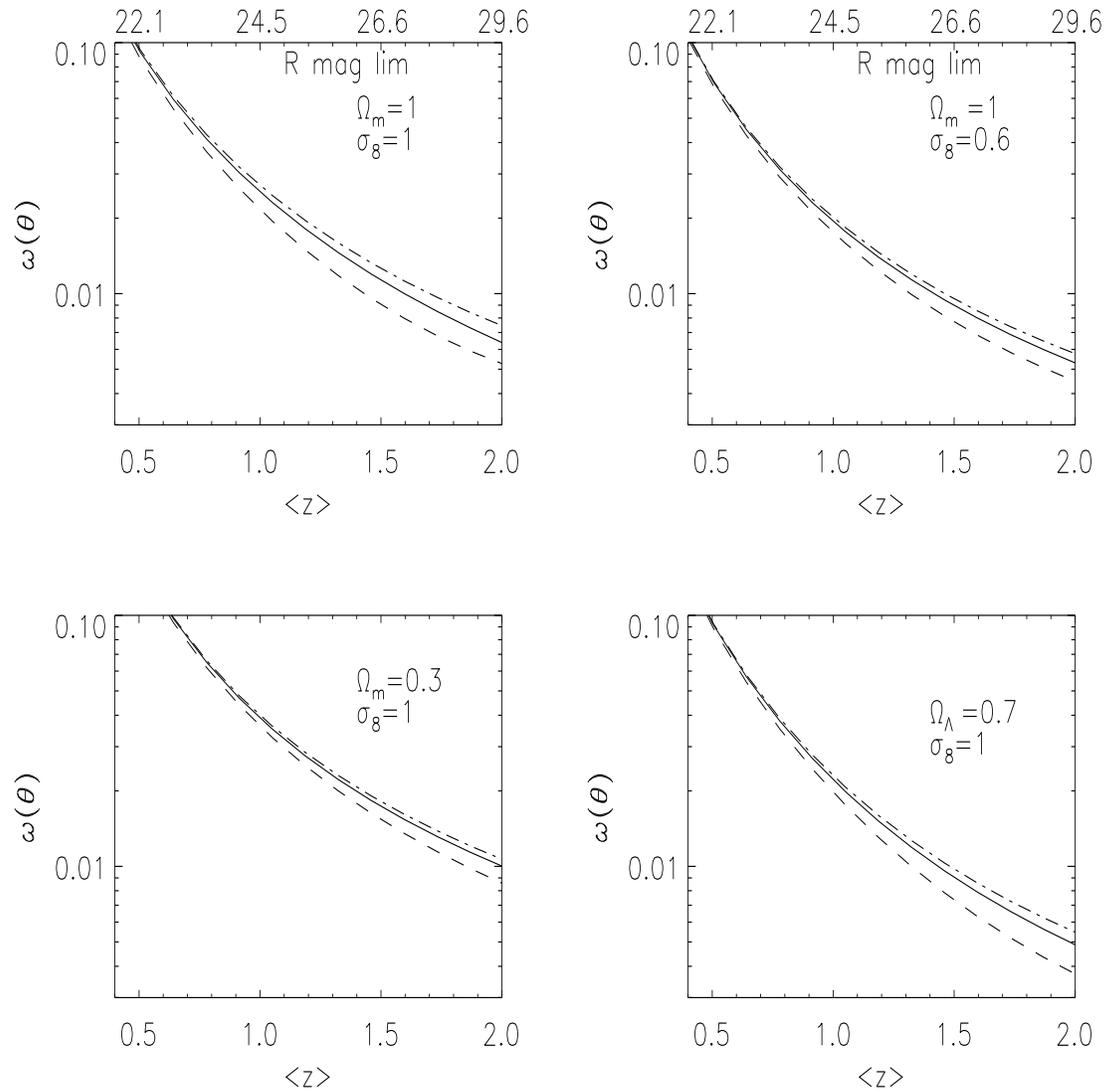,height=6in,width=6in}}
\spacing
 \caption{$\omega(\theta=1^\prime)$ as a function of mean redshift
          for $s=s_b$ (dot-dashed line), $s=0.4$ (solid line) and
	  $s=s_r$ (dashed line). The bias is $b=1/\sigma_8$, and the
	four cosmological models are as in Fig.~\ref{wintrinsic}.
         For the $\om=1$ models we also indicate the corresponding
         limiting $R$ magnitudes on the top axis.}
 \label{wauto2}
\end{figure}

In Fig.~\ref{wauto2}  we plot $\omega(\theta)$ as a
function of mean redshift for three values of
$s$, as in Fig.~\ref{wtheta}, and for all four cosmological models.
For the $\Omega_m=1$ models, we also give the $R$ magnitude limits
inferred from Charlot (1996) on the top axis of the plot,
corresponding to the mean redshifts shown on the bottom axis. These
values are given as an orientation only, since the relation between
$R$ magnitude and mean redshift is not
exactly linear, and depends on the model of galaxy
evolution used to derive it.
We can understand these results better by looking at the three
terms of $\wth$ in Eqs.~\ref{wterms} and ~\ref{wtilde} separately.
At low mean redshifts, the intrinsic clustering term $\omega_{gg}$
dominates, but it decreases with redshift. At low $\langle z \rangle$
the cross-term $\omega_{gl}$ is much smaller than the intrinsic term,
but it is approximately constant with redshift; $\omega_{ll}$ is even smaller
than the cross-term due to the
small lensing depth at small mean redshift, but it increases with  $\langle z
\rangle$.
As  the mean redshift of the sample increases,
$\omega_{gg}$
drops, and the cross-term starts to become noticeable. The cross-term
contains the factor $\Omega_m(s-0.4)$, so that for $(s-0.4) > 0$, $\wth$
is increased with respect to the intrinsic contribution due to
lensing, and for $(s-0.4) < 0$ it is
decreased.
Also, the redshift at which
this happens is larger for smaller $\Omega_m$ and $|s-0.4|$.
At still larger $\langle z \rangle$,  $\omega_{ll}$ has increased
sufficiently to become noticeable as well. Since
$\omega_{ll}\sim\Omega_m^2 (s-0.4)^2$ is always positive, its effect
goes in the same direction as that of the cross-term for  $(s-0.4)>
0$. But for $(s-0.4) < 0$, its effect is in the opposite
direction to that of the cross-term, so that the decrease due to $\omega_{gl}$
is partly compensated.

The flat model with the higher normalization $\sigma_8=1$ shows
a stronger lensing signal than the one with the lower $\sigma_8=0.6$,
because the bias factor $b=1/\sigma_8$ is smaller, which increases
the lensing contribution from the cross-term $\omega_{gl} \sim b$ relative
to the intrinsic clustering term $\omega_{gg} \sim b^2$.

In the open or $\Lambda-$ model, as compared to the flat model,
the change in growth rate
of perturbations and the spatial geometry affect
the intrinsic clustering part of $\wth$ as discussed above.
The geometric effect for $\Lambda>0$, which decreases the
intrinsic clustering, is the opposite of the lensing
part. Since the volume element is larger, the optical depth to
lensing is larger, and lensing is more effective. This explains why
the amplitudes for red and blue samples deviate more strongly than
in the $\om=1$ case in spite of the extra factor of $\om$ in front
of the cross-term $\tilde{w}_{gl}$.
For the open model also this geometric effect is present,
and compensates for some of the decrease due to the $\om$-prefactor,
but it is weaker than in the $\Lambda$ model.

\section{Comparison of red and blue samples}

The results presented so far for $\wth$ include the
contribution from intrinsic clustering and from lensing. It is
interesting to consider measurable statistics that are more sensitive
to the lensing contribution; in this section we show that the ratio
of $\wth$ for red and blue samples is one such example.
We assume the presence of a population of blue galaxies  with a number
count slope $s_b=0.45$ out to $z \largapprox 1$, and also of a population
of red galaxies  with a slope of $s_r=0.25$. A sample with such a
slope may be obtained by defining color selected subsamples \cite{VFC
96}, using the fact that the number count slope is a decreasing
function of $V-I$ color \cite{BVSC 97}.

Since red and blue galaxy samples are  observed
to have a relative bias at present, with red galaxies
being more strongly clustered (they are more likely to be found in the
centers of clusters) than blue ones, in this section
we consider a bias factor for the red sample $b_r$ which is larger than
the bias $b_b$ of the blue sample. We assume
the individual biases to be constant in time, and
we neglect possible physical evolution of the spatial
correlation function (besides the one due to gravitational evolution).

In the absence of lensing, and assuming one has selected samples with
the same redshift distribution, the ratio of $\wth$ of the blue and
red samples would be equal to the ratio of their relative biases
squared,
\begin{equation}
\frac{\omega(\theta;s_b)}{\omega(\theta;s_r)}=\left( \frac{b_b}{b_r}
\right)^2 \ .
\end{equation}
However, lensing changes this relationship.
In the regime where the cross-term dominates the third term, the
lensing contribution increases $\wth$ for the blue, and decreases it
for the red sample.

\begin{figure}
 \centerline{\psfig{figure=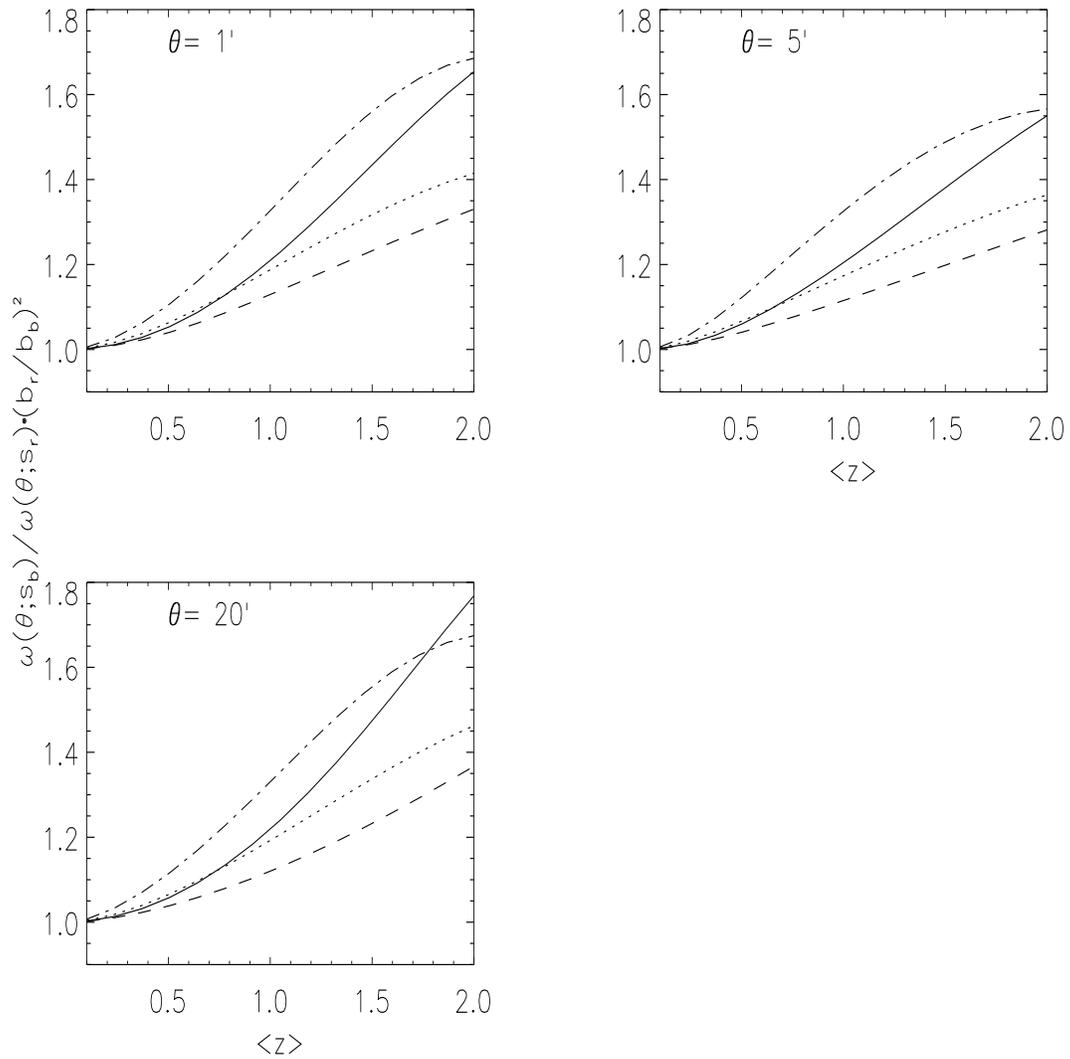,height=6in,width=6in}}
\spacing
 \caption{Ratio of $\wth$ for red and blue samples times the inverse
	  of the ratios of the bias factors squared
	  ($b=1/\sigma_8$, $b_r=b$,$b_b=b/2$).
	  Dash-dotted line: $\om=1$, $\sigma_8=1$; dotted line:
$\om=1$, $\sigma_8=0.6$; dashed line:  $\om=0.3$, $\sigma_8=1$; solid
line: $\om=0.3$, $\ol=0.7$,$\sigma_8=1$.
	  Three values of $\theta$ are shown.}
 \label{wratio}
\end{figure}

In Fig.~\ref{wratio} we plot $\left(b_r/b_b \right)^2
\omega(\theta;s_b)/ \omega(\theta;s_r)$, which ought to be  equal
to~$1$ in the absence of lensing, versus mean redshift for
$\theta=1^\prime, 5^\prime, 20^\prime$, and a relative bias of two of
the red and blue samples.
The four curves show the results for the different cosmological
models considered (see Section~2).
We can see that there is only little variation with angle as expected
from the results of Fig.~\ref{wtheta}.
For intermediate mean redshifts $\langle z \rangle \lessapprox 1$
we can neglect
the $\omega_{ll}$--term, and expand in the small quantity
$\tilde{\omega}_{gl}/ \tilde{\omega}_{gg}$ to see the dependence on
bias factors and number count slope,
\begin{eqnarray}
\frac{\omega(\theta;s_b)}{\omega(\theta;s_r)} & \approx &
\left( 1+ 30 \Omega_m \left[ \frac{s_b-0.4}{b_b}-\frac{s_r-0.4}{b_r} \right]
\frac{\tilde{w}_{gl}(\theta)}{\tilde{w}_{gg}(\theta)} \right)\nonumber \\
&& \times
\left( \frac{b_b}{b_r}\right)^2  \; .
\end{eqnarray}
For $\langle z \rangle \largapprox 1.5$,
$\omega_{ll}$
can no longer be neglected . As discussed in the
previous section, this term has the effect of compensating some of the decrease
in amplitude of
$\wth$ due to the cross-term for the red sample, and the exact redshift at
which this
happens depends on $\Omega_m$ and $(s-0.4)$. This effect is responsible for the
flattening off of  the
ratio seen in  Fig.~\ref{wratio}. At still larger redshifts, (not
plotted here), the ratio would decrease again.

As discussed in Section 3, for the open and $\Lambda-$dominated
models, both the growth rate
of perturbations and the spatial geometry are different from the
$\om=1$ model. By taking
the ratio of two correlation functions, the effect of the growth rate
cancels out, so that we are left with the effect of the different
geometries on the three terms.
For the $\Lambda-$dominated model the geometrical effect is largest,
so that in spite of the factor of $\om$ in the cross-term, the ratio
is larger than for the flat model normalized to
$\sigma_8=0.6$. It even attains a value comparable to the $\om=1$
model with the higher normalization of $\sigma_8=1$. Recall
that a larger $\sigma_8$ favors the lensing contribution over the
intrinsic clustering since the bias factor $b=1/\sigma_8$ is
smaller, and $\omega_{gl} \sim b$ whereas $\omega_{gg} \sim b^2$.
For the open model the
geometrical effect is not as important, and the ratio is below that
for the flat case.

\section{Conclusions}

We have quantified the effect of gravitational lensing by large-scale
structure on the angular correlation function of galaxies $\wth$, for
different cosmological models, on angular scales ranging
from $1^\prime$ to $20^\prime$. We have taken into account nonlinear
gravitational clustering which affects both the intrinsic clustering
and lensing contributions to $\wth$.

We find that the ratio of the angular correlation function for red and
blue galaxy samples, normalized by the inverse of the relative bias of
the two samples, deviates from the value of $1$ expected in the
absence of lensing at sufficiently large mean redshifts of the sample.
For a mean redshift of 1.3 of the sample, this ratio rises to about $1.5$ for
the model with $\om=1$ and $\sigma_8=1$, and continues to rise,
flattening off, until it reaches $1.7$ by a mean redshift of 2.
At $\langle z \rangle=1.3$ it rises to about $1.3$ for the
model with $\om=1$ and $\sigma_8=0.6$, and to about $1.2$ for the open model.
In the $\Lambda-$dominated model, the ratio reaches a value of about
$1.4$ at $\langle z \rangle=1.3$ , and continues to rise to
a value of about $1.6$ at a mean redshift of 2. These values are at $1^\prime$
angular separation; for larger angles they differ somewhat, but
not to a great extent as shown in Fig.~\ref{wratio}. The largest
uncertainty in the application of this result is due to the assumption of
two populations with different number count slopes but
the same redshift distribution with $\langle z \rangle \largapprox 1$.

The effect of magnification bias on the angular correlation function
could be used in future surveys, such as the ESO imaging survey
(Renzini et al. 1996),
to detect weak lensing by large-scale
structure and constrain the cosmological parameters $\om$ and $\ol$.
The ESO imaging survey aims to detect about 200-300 galaxies with
redshifts between $1$ and $2.8$ in a field of 25 arcminutes squared,
and a similar number with redshifts larger than $2.8$ in a field ten
times larger.
One problem with trying to detect
the effect of magnification bias on $\wth$ is that unknown physical
evolution of the galaxies can modify the intrinsic clustering
and introduce uncertainties which may be larger than the lensing
signal. The presence of bias evolution would introduce further
uncertainties.
In future work, we plan to study the
cross-correlation of two galaxy samples with different mean
redshifts and non-overlapping redshift distributions (Moessner \& Jain 1997).
In this way we hope to minimize the importance of
uncertainties due to intrinsic clustering and the exact form of
the redshift distribution, and propose a measurable
statistic that is dominated by the lensing contribution.

\section{Acknowledgements}
We would like to thank Matthias Bartelmann, Teresa Brainerd,
Peter Schneider, Uro\v s Seljak and Simon White for useful discussions.

\onecolumn

\end{document}